\pdfoutput=1
\documentclass{applemlr}

\usepackage{amsmath}
\usepackage{enumerate}
\usepackage{algorithm}
\usepackage{algpseudocode}
\usepackage{amsfonts}
\usepackage{amsthm}
\usepackage{cleveref}
\usepackage{diagbox}
\usepackage{colortbl}
\usepackage{amssymb}
\usepackage{xspace}
\usepackage{wrapfig}
\usepackage{adjustbox}
\usepackage{tabularx}
\usepackage{booktabs}
\usepackage{mathtools}
\usepackage{tikz}
\usepackage{enumitem}
\usepackage{silence}
\usepackage{dsfont}
\usepackage[table]{xcolor}
\usepackage[dvipsnames]{xcolor}
\usepackage{multirow}
\usepackage{makecell}
\usepackage{xfakebold}

\usepackage{amsmath,amsfonts,bm}

\def\eqref#1{equation~\ref{#1}}

\def\1{\bm{1}}

\DeclareMathAlphabet{\mathsfit}{\encodingdefault}{\sfdefault}{m}{sl}
\SetMathAlphabet{\mathsfit}{bold}{\encodingdefault}{\sfdefault}{bx}{n}

\definecolor{textgray}{HTML}{6E6E73}
\usetikzlibrary{positioning, calc}
\usetikzlibrary{decorations.pathmorphing}

\makeatletter
\patchcmd{\wrong@fontshape}{\@gobbletwo}{}{}{}
\makeatother
\numberwithin{equation}{section}
\makeatletter
\AtBeginDocument{
  \urlstyle{sf}
  
}
\makeatother

\definecolor{light}{RGB}{125, 125, 125}
\crefname{tcb@cnt@pbox}{code}{code}
\Crefname{tcb@cnt@pbox}{Code}{Code}
\crefname{assumption}{assumption}{assumption}
\Crefname{assumption}{Assumption}{Assumptions}

\newtcolorbox[auto counter]{pbox}[2][]{
  colback=white,
  title=Code~\thetcbcounter: #2,
  #1,fonttitle=\sffamily,
  fontupper=\sffamily,
  arc=2pt,
  colframe=bgcolor,
  coltitle=fgcolor,
  colbacktitle=bgcolor,
  toptitle=0.25cm,
  bottomtitle=0.125cm
}

\makeatletter
\newcommand\applefootnote[1]{%
  \begingroup
  \renewcommand\thefootnote{}%
  \renewcommand\@makefntext[1]{\noindent##1}%
  \footnote{#1}%
  \addtocounter{footnote}{-1}%
  \endgroup
}
\makeatother

\definecolor{cverbbg}{gray}{0.90}

\usepackage{ragged2e}
\usepackage{pifont}            
\usepackage{float}
\usepackage{threeparttable}
\usepackage{pgfplots}          
\pgfplotsset{compat=1.17}
\usepgfplotslibrary{groupplots}
\usetikzlibrary{arrows.meta, positioning, calc, backgrounds, fit}

\definecolor{cEB}{HTML}{154360}
\definecolor{cEF}{HTML}{EAF2FB}
\definecolor{cTB}{HTML}{145A32}
\definecolor{cTF}{HTML}{EAFAF1}
\definecolor{cDB}{HTML}{641E16}
\definecolor{cDF}{HTML}{FDEDEC}
\definecolor{cFg}{HTML}{1C1C1C}
\definecolor{cPanelBg}{HTML}{F5F6FA}
\definecolor{cPanelBd}{HTML}{CDD0D9}

\definecolor{oiBlue}{HTML}{0072B2}
\definecolor{oiVerm}{HTML}{D55E00}
\definecolor{oiGreen}{HTML}{009E73}
\definecolor{oiOran}{HTML}{E69F00}
\definecolor{oiPurple}{HTML}{CC79A7}
\pgfplotsset{
  ablbase/.style={
    width=\linewidth, height=5.0cm,
    axis line style={draw=gray!60, line width=0.6pt},
    tick align=outside, major tick length=3pt,
    tick style={gray!60, line width=0.5pt},
    xlabel={Training step}, ylabel={Training loss (EMA)},
    label style={font=\footnotesize}, tick label style={font=\scriptsize},
    xmajorgrids, ymajorgrids, scaled x ticks=false,
    grid style={draw=gray!22, line width=0.4pt},
    legend style={font=\scriptsize, draw=gray!45, line width=0.4pt,
                  fill=white, fill opacity=0.92, text opacity=1,
                  at={(0.985,0.985)}, anchor=north east, inner sep=3pt, row sep=1.0pt},
    legend cell align={left},
    every axis plot/.append style={line width=1.1pt, line cap=round, line join=round},
  },
  ablx25/.style={ablbase, xmin=0, xmax=25500, ymin=6.4, ymax=7.45,
    xtick={0,5000,10000,15000,20000,25000}, xticklabels={0,5k,10k,15k,20k,25k},
    ytick={6.4,6.6,6.8,7.0,7.2,7.4}},
  ablx100/.style={ablbase, xmin=0, xmax=102000, ymin=5.9, ymax=7.45,
    xtick={0,25000,50000,75000,100000}, xticklabels={0,25k,50k,75k,100k},
    ytick={6.0,6.2,6.4,6.6,6.8,7.0,7.2,7.4}},
  grpcommon/.style={
    width=\linewidth, height=3.35cm,
    axis line style={draw=gray!60, line width=0.6pt},
    tick align=outside, major tick length=2.5pt, tick style={gray!60, line width=0.5pt},
    label style={font=\scriptsize}, tick label style={font=\tiny},
    xmajorgrids, ymajorgrids, scaled x ticks=false, grid style={draw=gray!22, line width=0.4pt},
    legend style={font=\tiny, draw=gray!45, line width=0.4pt, fill=white, fill opacity=0.9,
                  at={(0.985,0.95)}, anchor=north east, inner sep=2pt, row sep=0.3pt},
    legend cell align={left},
    every axis plot/.append style={line width=0.9pt, line cap=round, line join=round},
  },
  grp25/.style={grpcommon, xmin=0, xmax=25500,
    xtick={0,5000,10000,15000,20000,25000}, xticklabels={0,5k,10k,15k,20k,25k}},
  grp100/.style={grpcommon, xmin=0, xmax=102000,
    xtick={0,25000,50000,75000,100000}, xticklabels={0,25k,50k,75k,100k}},
}

\tikzset{
  encblock/.style={draw=cEB, fill=cEF, rounded corners=6pt, line width=1.6pt,
    minimum width=3.0cm, minimum height=4.2cm, align=center},
  tdecblock/.style={draw=cTB, fill=cTF, rounded corners=6pt, line width=1.6pt,
    minimum width=3.0cm, minimum height=4.2cm, align=center},
  ddecblock/.style={draw=cDB, fill=cDF, rounded corners=6pt, line width=1.6pt,
    minimum width=3.0cm, minimum height=4.2cm, align=center},
  term/.style={draw=cFg!45, fill=white, rounded corners=4pt, line width=0.9pt,
    minimum width=2.0cm, minimum height=0.90cm, font=\small, align=center},
  opell/.style={draw=cFg!50, fill=gray!7, rounded corners=4pt, line width=0.9pt,
    minimum width=1.9cm, minimum height=0.78cm, font=\small, align=center},
  fwd/.style={-{Stealth[length=7pt, width=5pt]}, line width=1.4pt, draw=cFg},
  fwdlight/.style={-{Stealth[length=5pt, width=4pt]}, line width=0.85pt, draw=cFg!38, dashed},
  fbk/.style={-{Stealth[length=6pt, width=4.5pt]}, line width=1.15pt, draw=#1, rounded corners=6pt},
  fbk/.default=cFg!70,
  lbl/.style={font=\small, fill=white, inner sep=1.5pt, text=cFg}
}

\pgfdeclarelayer{background}
\pgfsetlayers{background,main}

\title{Memory Efficient Audio Synthesis with Decoupled Temporal Depth Diffusion Transformers}

\author[*]{Dongseong Hwang}
\author[*]{Prasanth Yadla}
\author[*]{Kaan Elgin}
\author{Shifas Padinjaru Veettil}
\author{Sivanand Achanta}
\author{Dipjyoti Paul}
\author{Ramya Rasipuram}
\author{Tyler Johnson}
\author{Emad Soroush}
\author{Chung-Cheng Chiu}
\author{Zhifeng Chen}

\affiliation{Apple}

\contribution[*]{These authors contributed equally to this work.}

\abstract{
Siri Expressive Voices synthesize rich, configurable speech in real time and entirely on device, powered by AFM~3 Core Advanced, Apple's most powerful on-device foundation model. This work presents the memory-efficient audio synthesis architecture behind that capability: a detokenizer that converts the semantic audio tokens emitted by the foundation model into high-fidelity audio within the tight compute and memory budget of the Scalable Matrix Extension (SME).

We convert semantic audio tokens to a residual vector quantization (RVQ) representation with a three-component design---a streaming encoder, a temporal decoder, and a depth decoder---that systematically decouples temporal and depth processing. A single reusable depth decoder with Diffusion Transformer (DiT)-style stage conditioning generates all RVQ levels autoregressively, replacing the dedicated per-level decoders of prior multi-decoder architectures, while causal sliding window attention with fixed-window key-value caching yields constant memory complexity independent of sequence length.

Deployed on SME, the detokenizer sustains roughly 10\,ms per generation step---about 16$\times$ faster than real time---with a peak runtime memory of only $\sim$21\,MB and 329\,MB of on-device assets, enabling continuous streaming synthesis of 20--320 seconds of audio alongside the on-device foundation model. This constant, small footprint replaces the linear and quadratic memory scaling of conventional transformer- and GAN-based approaches.

Comprehensive ablation studies validate the effectiveness of key architectural components, including DiT conditioning mechanisms, temporal lookahead processing, and unified depth decoding strategies. Audio quality assessment through phonetic discriminability analysis, perceptual quality metrics, and neural quality estimation confirms that the proposed architecture maintains synthesis fidelity while achieving computational efficiency gains over existing methodologies.

The proposed architecture is deployed in production as part of Siri Expressive Voices, powering a voice overhaul with Pace and Expressivity customization sliders in Apple Devices and support for custom assistant voices. Operating at a 1-billion-parameter activation size within AFM~3 Core Advanced, it improves Mean Opinion Score (MOS) by +0.28 overall (4.15 vs.\ 3.87) and by +0.42 on conversational speech (4.24 vs.\ 3.82) over the prior on-device text-to-speech system.
}

\metadata[Keywords]{\sffamily Speech Synthesis, Residual Vector Quantization, Diffusion Transformers, High-fidelity Synthesis}
\metadata[Correspondence]{\sffamily Dongseong Hwang: \url{dhwang2@apple.com}; Prasanth Yadla: \url{p_yadla@apple.com}}
\date{\sffamily\today}

\begin{document}

\maketitle

\section{Introduction}
\label{sec:intro}

The architecture presented here is deployed as the audio synthesizer for Siri Expressive Voices, a feature of AFM~3 Core Advanced, Apple's most capable on-device foundation model~\cite{apple2026afm3}. AFM~3 Core Advanced is natively multimodal and built on a sparse, 20-billion-parameter architecture that uses Instruction-Following Pruning (IFP)~\cite{hou2025instructionfollowingpruninglargelanguage}: the full model is stored in flash memory (NAND), and for each prompt a lightweight dense block selects a small subset of input-dependent \emph{routed} experts---1--4 billion active parameters, reselected periodically during generation---which combine with a set of always-active \emph{shared} experts to form a dense model in DRAM. Because only this small active set occupies DRAM at any time, audio generation must run within a tight, shared on-device memory and latency budget. Converting the foundation model's semantic audio tokens into high-fidelity audio under this budget is the challenge we address; we show that a memory-efficient detokenizer can perform the conversion on the Scalable Matrix Extension (SME) faster than real time and at a very small memory footprint, making streaming expressive synthesis practical on device.

Neural audio synthesis has evolved from raw waveform generation approaches like WaveNet \cite{oord2016wavenet} to more efficient architectures including WaveGlow \cite{prenger2019waveglow} and WaveRNN \cite{kalchbrenner2018efficient}. Transformer-based audio generation has been demonstrated in MusicLM \cite{agostinelli2023musiclm}, AudioLM \cite{borsos2023audiolm}, and Jukebox \cite{dhariwal2020jukebox}, though these approaches exhibit quadratic memory complexity limiting scalability to extended sequences. Diffusion-based methods, including DiffWave \cite{kong2020diffwave} and Diffusion Transformers (DiT) \cite{peebles2023scalable}, have shown promise but remain underexplored for real-time applications.

Contemporary speech generation uses discrete tokenization to exploit advances in large language models while providing computational efficiency for latency-critical applications \cite{défossez2022highfidelityneuralaudio} \cite{lakhotia2021generativespokenlanguagemodeling}. Vector quantization techniques, initially developed for image processing \cite{evans2024long}, have been successfully adapted to audio domains \cite{dhariwal2020jukebox}, with residual vector quantization (RVQ) \cite{lee2022autoregressive} addressing scalability limitations inherent in exponentially growing codebook sizes. While hierarchical RVQ representations enable extended audio generation capabilities, they necessitate multiple token predictions per timestep, creating computational bottlenecks. Semantic audio token representations \cite{kreuk2023audiogentextuallyguidedaudio} have emerged as a solution, providing single-token encodings of multi-layered RVQ structures.

Recent work by Moshi \cite{defossez2024moshi} demonstrates real-time conversational audio generation through a hierarchical RQ-Transformer that pairs a temporal transformer with a single Depth Transformer generating the RVQ levels autoregressively along the depth dimension. While the depth transformer body is shared across levels, Moshi retains codebook-specific parameters---in particular per-level input and output projections---so the effective parameterization still grows with the number of quantization levels.

To address computational limitations while maintaining generation quality, we propose a novel architecture for generating high-quality natural audio that decouples temporal and depth processing through specialized computational modules. Our approach generates quantized audio tokens from semantic audio token representations using a Diffusion Transformer (DiT)-style conditioning mechanism. While conceptually similar to Moshi's depth processing, our implementation achieves computational and parameter savings by employing a single \emph{fully} reusable depth decoder that identifies each RVQ level solely through DiT-style stage conditioning (a rotary encoding of the codebook index), eliminating even the per-level input and output projections used by Moshi. Concurrent work such as Fish Audio S2 \cite{fishaudio2026s2} adopts a related dual-autoregressive decoupling of temporal and depth processing, which further corroborates this design direction. A complementary line of work targets detokenizer efficiency from a different angle: T-Mimi \cite{wu2026tmimi}, building on TS3-Codec \cite{wu2024ts3codec}, replaces the de-convolution layers of Moshi's Mimi decoder with a purely transformer-based decoder to reduce mobile-CPU latency. In contrast, we address the upstream semantic-token-to-RVQ conversion and target the Scalable Matrix Extension (SME) rather than a mobile CPU.

The primary contributions of this work are: (1) a novel architecture that separates temporal and depth dimension processing for efficient high-quality audio generation, (2) a unified depth decoder design that reuses a single decoder with fully shared parameters across all RVQ levels through DiT-style stage conditioning, dispensing with the per-level projections retained by prior depth-transformer designs such as Moshi, (3) integration of a fixed-window KV cache mechanism that ensures constant memory consumption independent of audio sequence length, and (4) demonstration of superior computational efficiency and generation speed compared to state-of-the-art methods, including Moshi, enabling real-time natural audio generation in resource-constrained environments.

\section{Model Architecture}
\label{sec:pagestyle}

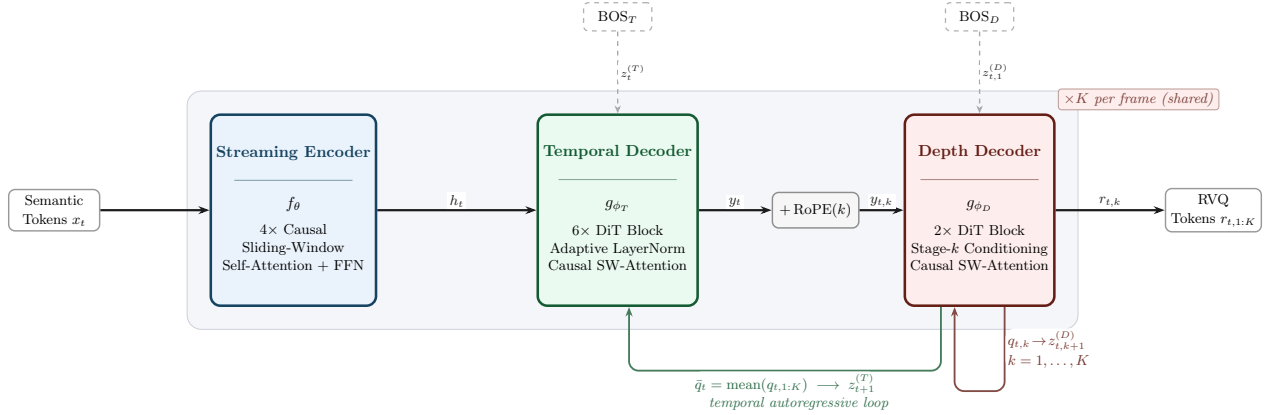
\begin{figure*}[t]
  \centering
  \resizebox{\textwidth}{!}{%
    \begin{tikzpicture}[node distance=0pt]
      \node[encblock] (ENC) at (0,0) {
        {\normalsize\bfseries\color{cEB}Streaming Encoder}\\[3pt]
        {\color{cEB!55}\rule{2.6cm}{0.35pt}}\\[5pt]
        {\normalsize $f_\theta$}\\[6pt]
        {\small\begin{tabular}{@{}c@{}}
          $4\times$ Causal\\[0pt] Sliding-Window\\[0pt] Self-Attention $+$ FFN
        \end{tabular}}
      };
      \node[tdecblock, right=3.5cm of ENC] (TDEC) {
        {\normalsize\bfseries\color{cTB}Temporal Decoder}\\[3pt]
        {\color{cTB!55}\rule{2.6cm}{0.35pt}}\\[5pt]
        {\normalsize $g_{\phi_T}$}\\[6pt]
        {\small\begin{tabular}{@{}c@{}}
          $6\times$ DiT Block\\[0pt] Adaptive LayerNorm\\[0pt] Causal SW-Attention
        \end{tabular}}
      };
      \node[ddecblock, right=4.5cm of TDEC] (DDEC) {
        {\normalsize\bfseries\color{cDB}Depth Decoder}\\[3pt]
        {\color{cDB!55}\rule{2.6cm}{0.35pt}}\\[5pt]
        {\normalsize $g_{\phi_D}$}\\[6pt]
        {\small\begin{tabular}{@{}c@{}}
          $2\times$ DiT Block\\[0pt] Stage-$k$ Conditioning\\[0pt] Causal SW-Attention
        \end{tabular}}
      };
      \node[term, left=2.4cm of ENC]   (SEM) {Semantic\\[1pt]Tokens $x_t$};
      \node[term, right=2.4cm of DDEC] (OUT) {RVQ\\[1pt]Tokens $r_{t,1:K}$};
      \node[opell, right=1.6cm of TDEC] (ROPE) {$+\,\mathrm{RoPE}(k)$};
      \node[draw=cFg!35, fill=white, rounded corners=3pt, line width=0.8pt, dashed,
        minimum width=1.5cm, minimum height=0.62cm, font=\small, align=center,
        above=1.8cm of TDEC] (BT) {$\mathrm{BOS}_T$};
      \node[draw=cFg!35, fill=white, rounded corners=3pt, line width=0.8pt, dashed,
        minimum width=1.5cm, minimum height=0.62cm, font=\small, align=center,
        above=1.8cm of DDEC] (BD) {$\mathrm{BOS}_D$};
      \begin{pgfonlayer}{background}
        \node[fill=cPanelBg, draw=cPanelBd, rounded corners=9pt, line width=0.55pt,
          fit=(ENC)(TDEC)(DDEC), inner sep=14pt] {};
      \end{pgfonlayer}
      \draw[fwd] (SEM)  -- (ENC);
      \draw[fwd] (ENC)  -- node[lbl, above]{$h_t$}     (TDEC);
      \draw[fwd] (TDEC) -- node[lbl, above]{$y_t$}     (ROPE);
      \draw[fwd] (ROPE) -- node[lbl, above]{$y_{t,k}$} (DDEC);
      \draw[fwd] (DDEC) -- node[lbl, above]{$r_{t,k}$} (OUT);
      \draw[fwdlight] (BT.south) -- node[lbl, right, font=\scriptsize]{$z_t^{(T)}$} (TDEC.north);
      \draw[fwdlight] (BD.south) -- node[lbl, right, font=\scriptsize]{$z_{t,1}^{(D)}$} (DDEC.north);
      \draw[fbk=cDB!80]
        ($(DDEC.south)+(+0.55, 0)$)
        -- node[lbl, right, text=cDB!90, font=\small, align=left]{%
            $q_{t,k}\!\to\!z_{t,k+1}^{(D)}$\\[1pt]$k=1,\ldots,K$}
        ($(DDEC.south)+(+0.55, -1.85)$)
        -- ($(DDEC.south)+(-0.55, -1.85)$)
        -- ($(DDEC.south)+(-0.55,  0)$);
      \node[fill=cDF, draw=cDB!45, rounded corners=2pt, line width=0.55pt,
        font=\small\itshape, text=cDB!85, inner xsep=3.5pt, inner ysep=2pt,
        anchor=south west] at ($(DDEC.north east)+(0.05, 0.07)$) {$\times K$ per frame (shared)};
      \draw[fbk=cTB!80]
        ($(DDEC.south)+(-0.85, 0)$)
        -- ($(DDEC.south)+(-0.85, -1.40)$)
        -- node[lbl, below, text=cTB!90]{%
            $\bar{q}_t = \mathrm{mean}(q_{t,1:K})\;\longrightarrow\;z_{t+1}^{(T)}$}
        ($(TDEC.south)+(+0.25, -1.40)$)
        -- ($(TDEC.south)+(+0.25,  0)$);
      \node[font=\small\itshape, text=cTB!85, anchor=north]
        at ($(TDEC.south)!0.5!(DDEC.south) + (0,-1.85)$) {temporal autoregressive loop};
    \end{tikzpicture}%
  }
  \caption{Proposed architecture comprising a streaming encoder $f_\theta$,
           temporal decoder $g_{\phi_T}$, and depth decoder $g_{\phi_D}$.
           The depth decoder is reused across all $K$ RVQ levels via
           DiT-style stage conditioning, with temporal feedback through
           the mean RVQ embedding $\bar{q}_t$.}
  \label{fig:arch}
\end{figure*}

As illustrated in Figure \ref{fig:arch}, the proposed architecture comprises three primary components: a streaming encoder, a temporal decoder, and a depth decoder.

The encoder functions as a feature extraction and representation learning module, transforming input semantic audio token representations into hierarchical condition latents that encapsulate temporal and semantic characteristics of the audio signal. This component establishes the mapping between diverse input token formats and the standardized internal representations required for subsequent processing stages.

The encoder employs causal sliding window attention \cite{beltagy20} to maintain causal constraints essential for autoregressive generation while managing computational complexity associated with extended audio sequences. Layer normalization \cite{ba2016layernormalization} is applied in a pre-norm configuration both preceding and following attention computations, enhancing training stability and convergence properties. Positional embeddings provide temporal information to guide the attention mechanism in processing sequential audio tokens, while residual connections around each attention block facilitate gradient propagation and enable deeper architectural configurations.

The temporal decoder functions as an intermediary module that bridges the encoder and depth decoder, processing condition latents from the encoder alongside previously generated RVQ tokens to predict subsequent token sequences. This component extracts essential information from the encoder's semantic audio token representations while incorporating feedback from generated RVQ tokens, ensuring that temporal dependencies are preserved throughout the generation process.

Crucially, the temporal decoder is required due to the RVQ autoregressive loop: each newly generated RVQ token must condition on previously generated tokens to maintain synthesis fidelity. Empirical analysis shows that removing the autoregressive loop leads to a severe drop in generated RVQ token accuracy, as the model loses access to critical temporal dependencies necessary for high-quality reconstruction.

This design contrasts with Moshi's approach, which simplifies the token mapping by utilizing identical input and output representations, whereas our architecture explicitly bridges semantic audio tokens and RVQ token sequences due to their representational differences and the need to preserve autoregressive dependencies.

The temporal decoder implements a Diffusion Transformer (DiT) \cite{peebles2023scalable} architecture incorporating causal sliding window attention and adaptive layer normalization modulation.

The depth decoder generates RVQ tokens autoregressively for each temporal frame using condition latents from the temporal decoder. Token generation for each frame is initialized using a learnable Beginning of Sequence (BOS) token without retaining previous states. Like Moshi, our depth decoder shares a single transformer body across all RVQ levels and generates them autoregressively along the depth dimension; unlike Moshi, it uses \emph{fully} shared parameters, identifying each level only through DiT-style stage conditioning rather than the per-level input and output projections Moshi employs.

The depth decoder architecture mirrors the temporal decoder, employing DiT \cite{peebles2023scalable} with causal sliding window attention and adaptive layer normalization modulation.

The architecture utilizes DiT-style conditioning, providing computational efficiency while maintaining effectiveness. This approach eliminates ambiguity in semantic audio to RVQ token mapping while avoiding computational overhead associated with cross-attention mechanisms.

Following equations describe the architecture. $x_t$ denotes the input token at timestep $t$, while $z_t^{(T)}$ and $z_t^{(D)}$ denote the time and depth decoder inputs respectively. Furthermore, $f_\theta$ denotes the encoder, while $g_{\phi_T}$ and $g_{\phi_D}$ denote the time and depth decoders. $K$ denotes number of codebooks.

\begin{align}
h_t &= f_\theta(x_t) \\
y_t &= g_{\phi_T}(z_t^{(T)}, h_t) \\
y_{t,k} &= \text{RoPE}(y_t, k) \\
(r_{t,k}, q_{t,k}) &= g_{\phi_D}(z_{t,k}^{(D)}, y_{t,k}) \\
z_{t,k+1}^{(D)} &= q_{t,k} \\
q_t &= \text{mean}(q_{t,1:K}) \\
z_{t+1}^{(T)} &= q_t
\end{align}

Time decoder conditions on the encoder output, while depth decoder conditions on the output of the time decoder. Time decoder output is augmented with the codebook index using rotary positional encoding before being fed as a condition information to the depth decoder. Depth decoder outputs RVQ tokens ($r_{t,k}$) and their corresponding embedding vectors ($q_{t,k}$). Finally, the mean of the embedding vectors is fed into the time decoder. The inputs of time and depth decoders are initialized with learnable BOS tokens. At every timestep, time decoder input is updated with the mean of the RVQ embedding vectors, while depth decoder input starts with a BOS token regardless of the timestep.

In addition, the architecture incorporates a temporal lookahead mechanism to enhance contextual understanding by providing the encoder with future semantic audio token information. This design enables the model to access contextual information beyond the current timestep, facilitating more informed generation decisions. Empirical analysis reveals that this temporal offset significantly impacts generated audio quality, indicating that enhanced contextual awareness through lookahead is crucial for achieving high-fidelity audio synthesis.

\section{Experiments}
\label{sec:experiments}

The transformer architectures incorporate contemporary architectural innovations derived from state-of-the-art language models, specifically LLaMA 3 \cite{metalama3blog} and Gemma 2 \cite{Google2024Gemma2}. The implementation utilizes RMS normalization \cite{zhang19} in conjunction with SwiGLU activation functions \cite{shazeer20}, rotary positional embeddings, and grouped query attention mechanisms to enhance computational efficiency and model expressiveness. Consistent with the design principles established in PaLM \cite{chowdhery2022palmscalinglanguagemodeling}, bias parameters are systematically excluded throughout the architectural framework. Logit scaling is implemented using a normalization factor of $\frac{1}{\sqrt{V}}$, where $V$ represents the vocabulary dimension. Hyperparameter optimization identified a learning rate of $1 \times 10^{-4}$ coupled with a cosine annealing schedule incorporating an extended peak phase as the configuration yielding superior convergence properties.

\subsection{On-Device Deployment}
\label{sec:ondevice}
The proposed architecture is deployed on SME as the audio detokenizer for Siri Expressive Voices in AFM~3 Core Advanced. The production deployment is tuned separately from the research configuration reported in this paper. To meet the stricter on-device compute and memory budget, the shipped model is further compressed and optimized, so it differs in several hyperparameters (e.g., layer allocation, number of RVQ levels, frame rate, and bitrate). Unless otherwise stated, the numbers reported in this paper refer to the research configuration. On device, it sustains a latency of approximately 10\,ms per generation step---producing 160\,ms of audio per step, roughly $16\times$ faster than real time (RTF~$\approx$~0.06)---with a peak runtime memory of only $\sim$21\,MB and 329\,MB of on-device assets (Table~\ref{tab:ondevice}). This small, constant footprint is what allows high-fidelity streaming synthesis to run concurrently with the sparsely-activated AFM~3 Core Advanced model under the device's shared memory budget. Because the fixed-window KV cache makes runtime memory independent of utterance length, the same budget supports everything from short Siri responses to long-form navigation and reading.

\begin{table*}[h!]
\centering
\caption{On-device performance of the proposed detokenizer on SME.}
\label{tab:ondevice}
\begin{tabular}{@{}lc@{}}
\toprule
\textbf{Metric} & \textbf{Value} \\
\midrule
Latency per generation step & $\sim$10\,ms \\
Throughput & 160\,ms audio / step ($16\times$ real time) \\
Peak runtime memory & $\sim$21\,MB \\
On-device assets & 329\,MB \\
\bottomrule
\end{tabular}
\end{table*}

\begin{table*}[h!]
\centering
\caption{Controlled cross-architecture comparison of algorithmic scaling across audio sequence lengths, measured on a common GPU so that all baselines are evaluated under identical conditions. The production deployment target is SME (Table~\ref{tab:ondevice}).}
\label{tab:performance_comparison}
\begin{tabular}{lcccc}
\toprule
\multirow{2}{*}{\textbf{Audio Length}} & \multirow{2}{*}{\textbf{Model}} & \textbf{Generation Time} & \textbf{Memory Usage} & \textbf{Real-Time Factor} \\
& & \textbf{(seconds)} & \textbf{(GB)} & \textbf{(RTF)} \\
\midrule
\multirow{3}{*}{512 tokens (20s)} & \textbf{Ours} & \textbf{2.40} & \textbf{1.13} & \textbf{0.12} \\
& Transformer Decoder & 0.78 & 1.78 & 0.04 \\
& Autoregressive GAN & 8.53 & 1.66 & 0.43 \\
\cmidrule(lr){1-5}
\multirow{3}{*}{2048 tokens (80s)} & \textbf{Ours} & \textbf{9.64} & \textbf{1.13} & \textbf{0.12} \\
& Transformer Decoder & 17.1 & 3.36 & 0.21 \\
& Autoregressive GAN & 33.97 & 1.66 & 0.42 \\
\cmidrule(lr){1-5}
\multirow{3}{*}{8192 tokens (320s)} & \textbf{Ours} & \textbf{38.59} & \textbf{1.13} & \textbf{0.12} \\
& Transformer Decoder & 800 & 33.44 & 2.5 \\
& Autoregressive GAN & 136.2 & 1.66 & 0.43 \\
\bottomrule
\end{tabular}
\begin{tablenotes}
\small
\item \textit{Note:} Reported on a single NVIDIA H100 GPU with unit batch size, used here only as a common platform for controlled comparison against baselines whose on-device numbers are unavailable. This isolates each architecture's scaling behavior: our model holds generation time, memory, and RTF constant with sequence length, whereas the baselines scale linearly or quadratically. Absolute on-device cost is far lower ($\sim$21\,MB peak runtime memory; see Table~\ref{tab:ondevice}).
\end{tablenotes}
\end{table*}

\begin{table*}[h!]
\centering
\caption{Ablation Study on architecture components. We evaluate semantic-to-RVQ token conversion quality by reporting evaluation loss. ``Temporal Lookahead'' refers to the number of future semantic tokens used for conditioning. ``Fixed KV Cache'' implements constant memory usage independent of sequence length.}
\label{tab:rvqformer_ablation}
\resizebox{\textwidth}{!}{%
\begin{tabular}{@{}ccccc|ccc@{}}
\toprule
\textbf{Temporal Lookahead} &
\textbf{Unified Depth Dec} &
\textbf{DiT Conditioning} &
\textbf{Output Norm} &
\textbf{Fixed KV Cache} &
\textbf{Eval Loss ($\downarrow$)} \\
\midrule
\ding{55} & \ding{55} & \ding{55} & \ding{55} & \ding{55} & 5.86 \\
\ding{51} & \ding{55} & \ding{55} & \ding{55} & \ding{55} & 4.09 \\
\ding{51} & \ding{51} & \ding{55} & \ding{55} & \ding{55} & 3.34 \\
\ding{51} & \ding{51} & \ding{51} & \ding{55} & \ding{55} & 3.28 \\
\ding{51} & \ding{51} & \ding{51} & \ding{51} & \ding{55} & 3.23 \\
\ding{51} & \ding{51} & \ding{51} & \ding{51} & \ding{51} & \textbf{3.23} \\
\bottomrule
\end{tabular}%
}
\end{table*}

The model underwent pretraining on an extensive proprietary audio corpus characterized by substantial dataset diversity. The training data is comprised of many hours of internally recorded  voice assistant interactions. Detailed dataset compositions and specifications cannot be disclosed in accordance with institutional data governance policies and confidentiality protocols.

Beyond the on-device deployment numbers in Section~\ref{sec:ondevice}, we compare algorithmic scaling against baselines on a common GPU, since on-device measurements are not available for the external baselines. Table~\ref{tab:performance_comparison} reports generation time, memory, and real-time factor (RTF) as a function of sequence length. The proposed architecture holds all three approximately constant with sequence length, whereas the transformer-decoder and autoregressive-GAN baselines scale linearly or quadratically. This constant-scaling property is what carries over to the device: as shown in Table~\ref{tab:ondevice}, the deployed detokenizer runs at $16\times$ real time using only $\sim$21\,MB of runtime memory, independent of utterance length.
Additionally we conducted several ablation studies to determine the best configuration for training the model which are described in Table \ref{tab:rvqformer_ablation}.

\begin{table*}[h!]
\centering
\caption{Audio Quality Evaluation of Neural Audio Codecs. All entries use the research configuration for controlled comparison; the shipped production configuration differs.}
\label{tab:audio_quality}
\begin{tabular}{@{}lcccccccc@{}}
\toprule
\textbf{Model} & \textbf{$f_s$} & \textbf{$f_r$} & \textbf{Bitrate} & \textbf{UTMOS ($\uparrow$)} & \textbf{SI-SNR ($\uparrow$)} \\
\midrule
Ground Truth        & 24kHz & -- & -- & 4.07 & -- \\
Moshi/Mimi          & 24kHz & 12.5Hz & 1.1kbps & 3.92 & 7.45 \\
Transformer Decoder & 24kHz & 25Hz & 1.5kbps & 3.61 & 3.76 \\
Autoregressive GAN  & 24kHz & 25Hz & 1.5kbps & 3.91 & 5.50 \\
\textbf{Ours}       & 24kHz & 25Hz & 1.5kbps & 3.97 & 9.47 \\
\bottomrule
\end{tabular}
\end{table*}

\subsection{Conditioning Method}

\begin{figure}[htbp]
    \centering
    \begin{subfigure}[t]{0.48\linewidth}
        \centering
        \includegraphics[width=\linewidth]{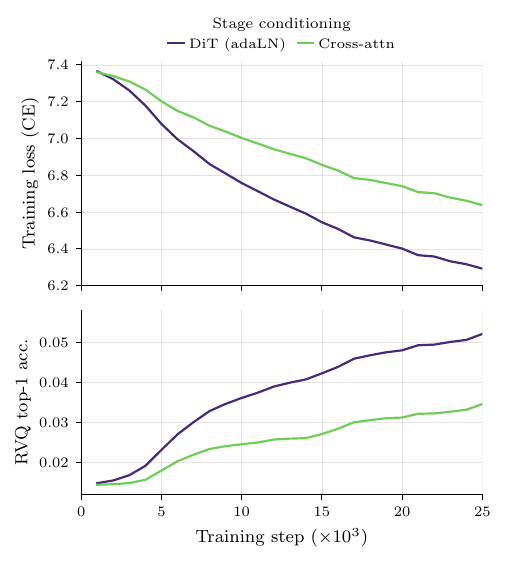}
        \caption{DiT-style conditioning versus cross-attention.}
        \label{fig:dit}
    \end{subfigure}\hfill
    \begin{subfigure}[t]{0.48\linewidth}
        \centering
        \includegraphics[width=\linewidth]{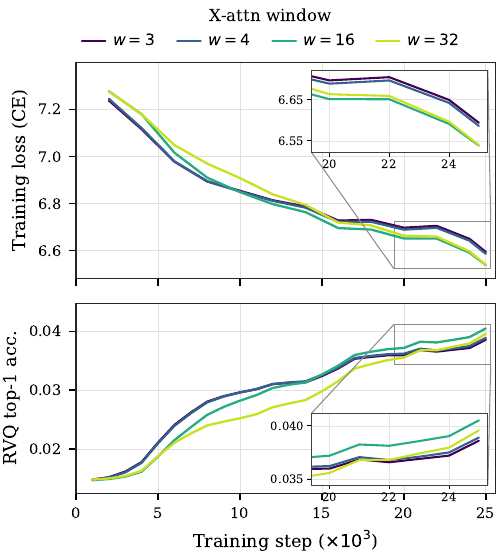}
        \caption{Cross-attention across context lengths. With gradient clipping and a tuned learning rate, all lengths---including large windows ($\geq32$) that previously diverged---train stably.}
        \label{fig:xattn}
    \end{subfigure}
    \caption{Conditioning-mechanism ablations: training loss for the two conditioning schemes (left) and for cross-attention context-window sizes (right).}
    \label{fig:conditioning}
\end{figure}

Comparative analysis of conditioning mechanisms shows that DiT-style conditioning matches cross-attention in generation quality while being substantially more efficient. On held-out evaluation datasets, the two are statistically indistinguishable in both loss and next-token accuracy, yet DiT-style conditioning trains 1.36$\times$ faster and avoids the additional attention computation that cross-attention requires over the conditioning context. It also removes a hyperparameter --- the cross-attention context window --- that we found had no consistent effect on performance. These properties make DiT-style conditioning the preferred choice for the proposed architecture. Figure~\ref{fig:dit} compares the training-loss trajectories of the
two mechanisms.

Further investigation of cross-attention performance shows that the context-window size has no consistent effect on quality. Earlier experiments exhibited divergent training loss at extended context lengths (32 or greater); we found this to be a training-stability artifact rather than an architectural limitation: with gradient clipping and a tuned learning rate, large cross-attention context lengths train stably without divergence. Cross-attention thus introduces an additional hyperparameter, the context window, without providing a consistent quality benefit, further motivating the choice of DiT-style conditioning. Figure~\ref{fig:xattn} presents the training loss characteristics across different cross-attention configurations under this stabilized training setup.

\subsection{Time Lookahead}
\begin{figure}[htbp]
    \centering
    \begin{subfigure}[t]{0.48\linewidth}
        \centering
        \includegraphics[width=\linewidth]{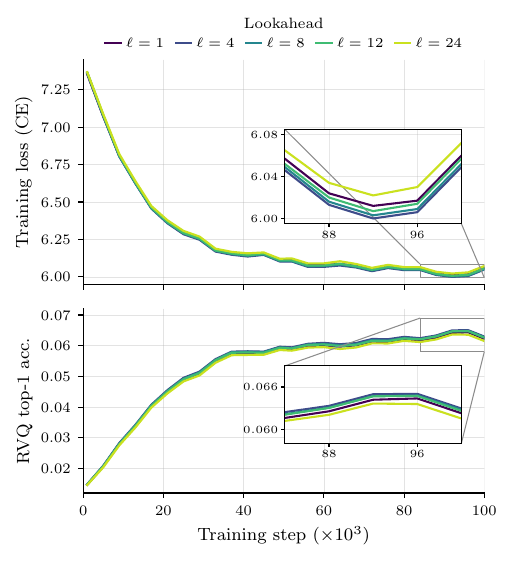}
        \caption{Varying temporal lookahead configurations.}
        \label{fig:lookahead}
    \end{subfigure}\hfill
    \begin{subfigure}[t]{0.48\linewidth}
        \centering
        \includegraphics[width=\linewidth]{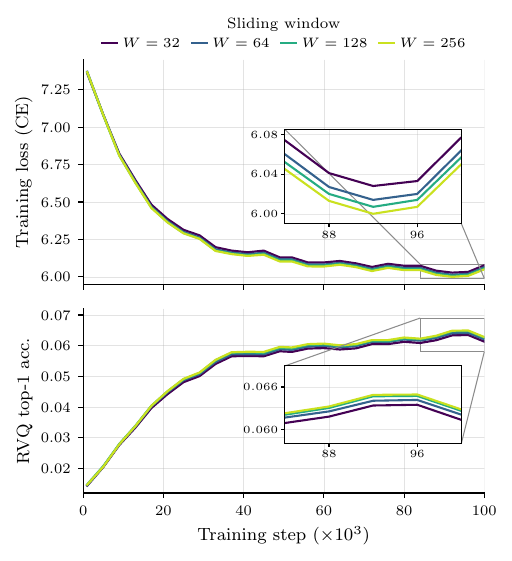}
        \caption{Varying sliding window sizes.}
        \label{fig:window}
    \end{subfigure}
    \caption{Attention-context ablations: training loss across temporal lookahead settings (left) and sliding window sizes (right).}
    \label{fig:context_ablations}
\end{figure}

Temporal lookahead analysis reveals architectural differences compared to existing approaches. While Moshi's experimental validation established that a single-frame lookahead suffices for their architecture, the proposed framework requires a 6-token temporal offset to achieve optimal performance. Comparative evaluation of bidirectional encoding demonstrates marginal accuracy improvements, suggesting that the 240ms lookahead window introduces negligible information loss from the input signal while providing sufficient contextual information for high-quality generation. Figure~\ref{fig:lookahead} presents the training loss characteristics across different temporal lookahead configurations.

\subsection{Sliding Window Size}
Sliding window size optimization identified 128 tokens as the configuration yielding optimal performance characteristics. The initial hypothesis favored a constrained window size, predicated on the assumption that the model, operating as a deterministic token-to-token mapping function, would not require access to distant temporal context for effective generation. However, empirical analysis revealed that the model possesses sufficient representational capacity to selectively attend to salient information even when provided with extended contextual windows. These findings suggest that larger window sizes do not introduce detrimental effects on model performance, indicating robust attention mechanisms capable of filtering relevant information from extended sequences. Figure~\ref{fig:window} illustrates the training loss behavior across different sliding window configurations.

\subsection{Layer depth ratio analysis}
Architectural depth optimization analysis reveals that a layer configuration of 4-6-2 (encoder-temporal decoder-depth decoder) achieves optimal performance across loss, evaluation accuracy, and training efficiency metrics. The depth decoder exhibits the highest computational cost due to its iterative execution across all quantization levels, requiring six iterations per token, yet performance saturates at a depth of two layers. Configuration with zero layers yields substantially degraded performance, as it attempts to predict all RVQ tokens simultaneously, indicating that a minimum of one layer is essential for effective hierarchical token generation.

\begin{figure}[htbp]
    \centering
    \begin{subfigure}[t]{0.48\linewidth}
        \centering
        \includegraphics[width=\linewidth]{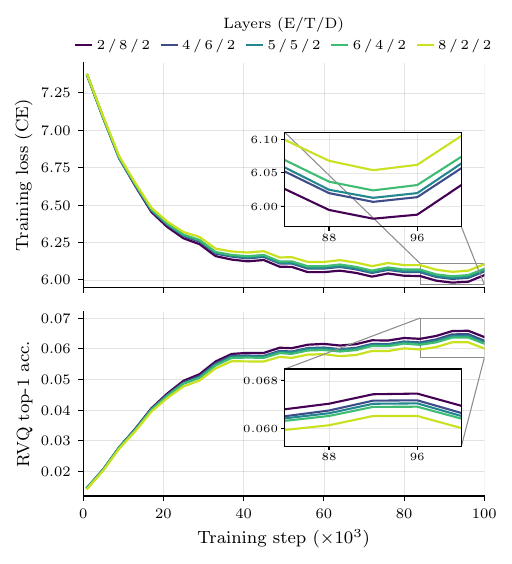}
        \caption{Varying layer depth ratios for the encoder, temporal decoder, and depth decoder.}
        \label{fig:layer_depth_ratio}
    \end{subfigure}\hfill
    \begin{subfigure}[t]{0.48\linewidth}
        \centering
        \includegraphics[width=\linewidth]{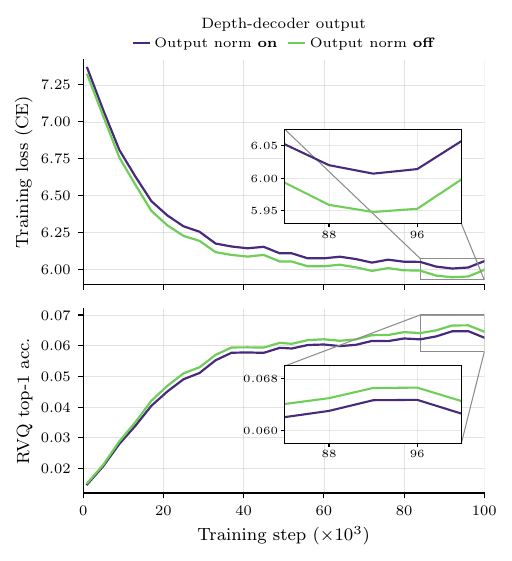}
        \caption{With and without output normalization.}
        \label{fig:output_norm}
    \end{subfigure}
    \caption{Training loss for the layer-depth-ratio ablation (left) and the output-normalization ablation (right).}
    \label{fig:depth_norm_ablations}
\end{figure}

With the depth decoder configuration fixed at two layers, systematic optimization of encoder and temporal decoder depths was conducted. Results demonstrate that four layers for the encoder and six layers for the temporal decoder provide superior performance across all evaluated metrics, including training loss, computational efficiency, and evaluation accuracy. The increased depth requirement for the temporal decoder is attributed to its critical role in aggregating comprehensive information from both encoder outputs and previously generated tokens to produce effective latent representations for subsequent depth processing. Figure~\ref{fig:layer_depth_ratio} illustrates the training loss behavior across different layer depth configurations.

\subsection{Output Normalization Analysis}

Following established practices in text-based large language models, output normalization is applied immediately preceding logit computation. Comparative analysis reveals that while omitting normalization yields marginally superior performance during early training phases, models with normalization demonstrate enhanced convergence stability and superior long-term performance. This aligns with language modeling literature, where output normalization serves as a regularization mechanism improving training dynamics. Figure~\ref{fig:output_norm} presents training loss trajectories comparing models with and without output normalization.

\subsection{Audio Quality Evaluation}

One way to measure the effectiveness of the compression method is to measure the phonetic discriminability of the representation space, which means comparing how well each approach preserves phonetic distinctions after reconstruction. Inspired from Moshi, we compute the within-speaker ABX \cite{schatz13_interspeech} metrics on LibriHeavy \cite{Panayotov2015LibrispeechAA} and compare the error rates of our model with three different baseline models, which are Mimi decoder (Moshi's decoder), an internal transformer decoder, and another internal model which is similar to the autoregressive GAN approach \cite{morrison2022chunkedautoregressiveganconditional}. We measure error rates of the quantized vector outputs from our model, formed by the sum of RVQ embeddings at each layer. Our representation achieves a within-speaker ABX error rate of 5.3\%, confirming that phonetic distinctions are well preserved after reconstruction.

Furthermore, in order to measure acoustic quality of reconstructed audio, we use VisQOL \cite{Hines2015ViSQOLAO}, MOSNet \cite{Lo2019MOSNetDL}, and SI-SNR metrics.

\subsection{Production Human Evaluation}

The expressive voices powered by AFM~3 Core Advanced were evaluated against Apple's prior production text-to-speech system through crowd-sourced Mean Opinion Score (MOS) evaluations across five production voices spanning three domains: Siri (task-oriented assistant speech), Conversational (natural dialogue-style speech), and Navigation (turn-by-turn direction speech).

Operating at a 1-billion-parameter activation size, AFM~3 Core Advanced achieves an overall MOS of 4.15 versus 3.87 for the production baseline---a +0.28 improvement, notable given that a 0.1 MOS gain is already highly perceptible to users. The gain is largest on conversational speech, where AFM~3 Core Advanced scores 4.24 versus 3.82 (+0.42), reflecting markedly more natural and expressive cadence. Audio cleanliness---the fraction of samples free of tremor, slurs, and glitches---also improves, with Navigation reaching 100\% clean, confirming a systematic reduction of perceptible artifacts in production.

\section{Conclusions}
\label{sec:conclusion}

This work establishes a novel paradigm in neural speech synthesis through the introduction of a semantic audio token processing architecture that fundamentally diverges from existing approaches such as Moshi. While Moshi generates the RVQ levels with a single depth transformer that retains codebook-specific input and output projections, our three-component architectural design achieves superior computational and parameter efficiency through a single reusable depth decoder whose parameters are fully shared across levels via DiT-style stage conditioning. The proposed system generates high-quality natural audio from semantic audio token inputs via this specialized neural architecture that decouples temporal and depth processing, eliminating the per-level parameterization overhead of Moshi's depth transformer. The integration of diffusion transformer based conditioning mechanisms with causal sliding window attention and fixed-window KV cache enables enhanced synthesis quality while maintaining computational efficiency required for real-time deployment. Deployed on SME as the audio synthesizer for Siri Expressive Voices in AFM~3 Core Advanced, the architecture runs at roughly $16\times$ real time ($\sim$10\,ms per generation step) using only $\sim$21\,MB of runtime memory and maintains constant memory across audio sequences ranging from 20 to 320 seconds---in contrast to Moshi and other state-of-the-art methodologies that exhibit linear or quadratic memory scaling. This memory efficiency is what enables high-quality, expressive, real-time speech synthesis to run entirely on device alongside a sparsely-activated foundation model, addressing fundamental scalability limitations in contemporary audio generation systems.

\bibliographystyle{plainnat}
\bibliography{refs}

\applefootnote{ \textcolor{textgray}{\sffamily Apple and the Apple logo are trademarks of Apple Inc., registered in the U.S. and other countries and regions.}}

\end{document}